# Identification of Nonlinear Systems From the Knowledge Around Different Operating Conditions: A Feed-Forward Multi-Layer ANN Based Approach

Sayan Saha[1], Saptarshi Das[2], Anish Acharya[1]

1. Department of Instrumentation and Electronics Engineering, Jadavpur University, Salt-Lake Campus, LB-8, Sector 3, Kolkata-700098, India.
2. Department of Power Engineering, Jadavpur University, Salt-Lake Campus, LB-8, Sector 3, Kolkata-700098, India.
Email: saptarshi@pe.jusl.ac.in, s.das@soton.ac.uk

Abhishek Kumar[2,3], Sumit Mukherjee[2,4], Indranil Pan[2], Amitava Gupta[2]

3. Electrical Engineering Department, National Institute of Technology-Hamirpur, Hamirpur-177005, HP, India.
4. Department of Electrical, Computer and Systems Engineering, Rensselaer Polytechnic Institute, Troy, New York-12180, United States.

*Abstract*—The paper investigates nonlinear system identification using system output data at various linearized operating points. A feed-forward multi-layer Artificial Neural Network (ANN) based approach is used for this purpose and tested for two target applications i.e. nuclear reactor power level monitoring and an AC servo position control system. Various configurations of ANN using different activation functions, number of hidden layers and neurons in each layer are trained and tested to find out the best configuration. The training is carried out multiple times to check for consistency and the mean and standard deviation of the root mean square errors (RMSE) are reported for each configuration.

## I. INTRODUCTION

Mathematical modeling of real life processes has got great importance for better understanding of the system's underlying physics and to predict or simulate its dynamical behavior. Modeling of a physical system can be done from the knowledge of system's excitation and corresponding response which is commonly known as system identification in the control engineering community [1]. It is fact that most of the physical systems are nonlinear though for controller design problems linear model development has been the most popular means. Contemporary researchers have given a brief overview of various nonlinear system identification techniques including neuro-fuzzy approaches [2]-[4]. A system can be identified both on the basis of polynomial based static model approach and regression based dynamic model approach [2]. However, we are interested in deriving the dynamic model instead of the static one (by means of lookup table or curve-fitting techniques) since the dynamic model can be written in terms of difference equation or differential equation in discrete or continuous time respectively. Though for most real systems, the responses are nonlinear functions of the input excitations, such system can be modeled accurately by a linear model in most cases. However, for the cases where the governing equation of the physical phenomenon is inherently nonlinear or the linearized model is not accurate, nonlinear system modeling is preferred. The classical methods used for identifying a dynamic model of a nonlinear system are Nonlinear Auto Regressive (NARX) models and Hammerstein-Wiener models [1]. NARX models use a finite number of previous and present input values and past output values to predict the present output of the system. Hammerstein-Wiener class of models is a series combination of static nonlinear blocks (one on each side of the linear block or only on one side of it), which captures the inherent nonlinearity of the system and a dynamic linear block to capture the system's dynamics. A more recent approach is to use Adaptive Neuro-Fuzzy Inference System (ANFIS), where the fuzzy inference rules are tuned with a neural network model to fit the given set of input and output data [5]. Another neuro-fuzzy system called Local Linear Model Tree (LOLIMOT) which is a type of Takagi-Sugeno-Kang neuro-fuzzy algorithm, is basically an adaptive network and provides robust learning capabilities. This algorithm divides the input space into local linear models which has a higher performance and needs lower neuron count compared to normal neural networks in terms of learning a mapping of the input space.

One major shortcoming of the methods, considered above is that none of them is capable of modeling the real system based on the information about its operating points. If we have a nonlinear differential equation characterizing a dynamical system and knowledge about different operating points of the system, then a linearized process model of the system can be obtained by expanding the Taylor Series of the governing nonlinear differential equation about each of these operating points. However, the obtained models would differ significantly if different operating points are used for the series expansion. Instead, here the proposed method obtains a single nonlinear dynamical model of the system, which replicates the system accurately, capturing its nonlinearity by incorporating the information around these operating points. An Artificial Neural Network, trained by updating its weights interconnecting different neurons, is used as a single nonlinear model of the original system using available information around different operating points i.e. the local linear behaviors.

To demonstrate the efficacy of the proposed method two inherently nonlinear systems have been considered, the nuclear reactor power level monitoring system and the AC servo position control system. In case of a nuclear reactor the output power level varies in a nonlinear fashion with the initial power of the reactor expressed as percentage of the full power of the reactor and the amount of inserted negative reactivity which is expressed in terms of the length of control rod dropped into the reactor. The nonlinearity in nuclear reactor is due to the point-kinetic governing equation having cross product of state variables and external excitation (input). On the other hand, the nonlinearities inherent in the AC servo position control system are due to the friction, hysteresis, back-lash etc. [6]. For both of these systems, input-output datasets consisting local linear behaviors around different operating points within the normal operating range were recorded. This was utilized to train the feed-forward multilayer artificial neural networks to obtain a nonlinear dynamic model which is capable of capturing all the local dynamics around each of the operating points; as a whole to represent the actual nonlinear system accurately [7]-[8].

Rest of the paper is organized as follows. Section II describes the basics of ANN and its application in modeling of nonlinear systems. A brief description of two target applications i.e. the nuclear reactor power level monitoring system and the AC servo position control system is presented in Section III. Identification and model simulation results for both of the systems are outlined in Section IV. The paper ends with the conclusion as Section V, followed by the references.

## II. BASICS OF THE FEED-FORWARD MULTI-LAYER ARTIFICIAL NEURAL NETWORK

Artificial Neural Network, consisting of a number of interconnected artificial neurons with linear or non-linear transfer (activation) functions is a tool that operates by mimicking the neural structure of the human brain. It is capable of capturing and predicting non-linear behavior of a system. ANNs have been widely used in the field of control systems for the purpose of system identification, nonlinear modeling, gain-adaptation etc. [9]-[10]. Other ANN based applications like identification in noisy environment [11], prediction [12] and controller parameter scheduling [13] have also been popular. An ANN predicts the system behavior by learning the relations between inputs to the system and the corresponding outputs of the system through an iterative training process. Its wide applicability stems from the fact that the weights of the individual neurons can be trained with several optimization algorithms (like gradient based methods, Levenberg-Marquardt back-propagation, evolutionary and swarm algorithms etc. [14]) so that the entire network can ultimately approximate almost any given non-linear function. There are no specific guidelines for choosing the number of hidden layers, bias weights, choice of interconnections, activation functions etc. in a specific neural network architecture and mostly depend upon the users intuition [7]-[8].

The multi-layer feed-forward network is the type of network used during this study. It consists of an input layer, one or two hidden layers and an output layer as shown in Fig. 1. The back-propagation algorithm which is used to train the neural network consists of two phases i.e. the forward phase and the backward phase. Assuming a training data, being defined as $[x(n), d(n)]$, with the input vector $x(n)$ applied to the nodes of the input layer and $d(n)$ being the corresponding desired output vector. During the forward phase, for each layer the local induced field for each neuron is then determined as $v_j^l(n) = \sum_{i=0}^{m_0} w_{ji}^l(n) y_i^{l-1}(n)$ ; where, $m_0$ is one less than the number of neurons in previous layer, $v_j^l(n)$ is the induced local field for neuron $j$ in the layer $l$, $y_i^{l-1}(n)$ is the output signal of neuron $i$ in the previous layer $l-1$ at the iteration $n$ and $w_{ji}^{l-1}(n)$ is the synaptic weight connecting neuron $i$ in the previous layer $l-1$ with the neuron $j$ in the layer. Activation functions, like, hyperbolic sigmoid (*tansig*) and logarithmic sigmoid (*logsig*) are then used to obtain the output of each node as $y_j^l = \phi_j(v_j(n))$, where $\phi_j(\cdot)$ is the activation function.

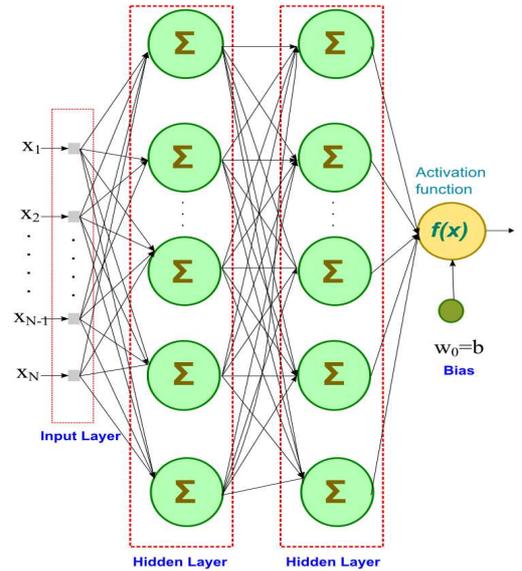

Figure 1. Schematic diagram of a feed-forward multilayer ANN.

In case the neuron $j$ is in the first hidden layer, set $y_j^0(n) = x_j(n)$, the $j^{th}$ element of the input vector $x(n)$. If the neuron is in the output layer then set $y_j^L = o_j(n)$, the $j^{th}$ element of the obtained output vector $o(n)$. The error signal $e_j(n)$ is the difference of the desired response and the obtained response. Synaptic weights of the network in layer $l$ are then updated in the backward pass as:

$w_{ji}^l(n+1) = w_{ji}^l(n) + \alpha[w_{ji}^l(n-1)] + \eta \delta_j^l(n) y_i^{l-1}(n)$, where,

$$\delta_j^l(n) = \begin{bmatrix} e_j^L(n)\phi_j'(v_j^L(n)) & \text{for neuron j in output layer L} \\ \phi_j'(v_j^l(n))\sum_{k=0}^{m_0}\delta_k^{(l+1)}(n)w_{kj}^{(l+1)}(n) & \text{for neuron j in hidden layer l} \end{bmatrix}$$

and $\eta$ is the learning rate parameter and $\alpha$ is the momentum constant of the algorithm. After completion of training in this way the weights are fixed and the network can now be used to predict accurately the output corresponding to any unknown input value provided it is within the range of the input values used to train the network.

For the nuclear reactor power level monitoring system, the inputs to the multilayer feed-forward neural network are the instantaneous position of the control rod (in fraction of the full length), time elapsed (in seconds), initial power (in percentage), and percentage of rod drop (30% or %50%) as in [15] respectively. The output of the system is the instantaneous power of the nuclear reactor. Similarly the inputs to the neural network predicting the behavior of the AC servo motor position control system are the time elapsed (in seconds), acceleration, velocity and output is the instantaneous position.

### III. TARGET APPLICATIONS WITH THE PROPOSED NONLINEAR IDENTIFICATION METHODOLOGY

#### A. Nuclear Reactor Power Level Monitoring System

For nonlinear system identification, a nuclear reactor is visualized as a system with control rod position (fraction of total drop) as input and the global power (in percentage of maximum power produced) as the output. The identification is based on data obtained from operating Indian PHWRs provided by Nuclear Power Corporation of India Ltd. (NPCIL) as also studied in Das *et al.* [15]. The data at different step-back levels is provided for 14 seconds with 0.1 second of sampling time. Graphical representation of the data is shown in Fig. 2 for 30% and 50% rod drop cases with different initial powers i.e. 100%, 90%, 80% and 70%. The dynamics of nuclear reactors, governed by nonlinear point kinetic equation is thus heavily dependent on the initial value of the state variables (operating reactor power) and the external excitation (negative reactivity or equivalent control rod worth) [16]. Basically, ANN has been employed here to identify the reactor from the 30% and 50% rod-drop data while the reactor is operating at various power levels.

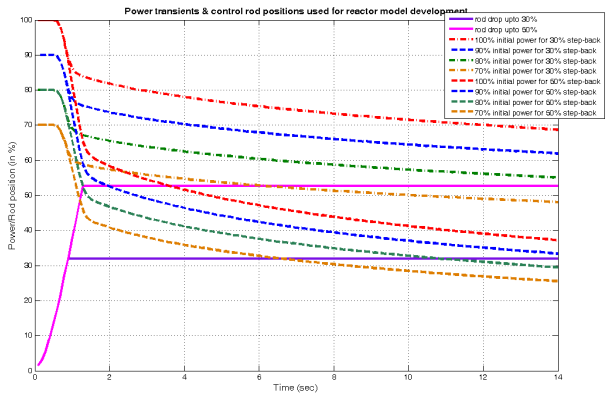

Figure 2.  Power transients and rod drop data used for reactor modeling.

#### B. AC Servo Position Control System

A microcomputer controlled AC servo system consisting of Versa Motion Servomotor and Versa Motion Servo-drive [17], manufactured by GE Fanuc (Fig. 3) is used in this study. The servo drive can itself control the motor (as the gain of controller can be set manually in the drive) or can be connected to external controller to control the motor. The servo motor has an inbuilt incremental encoder that gives 2500 pulse per revolution (10,000 quadrature count per revolution) resulting in a resolution of 0.036 degree angular rotation.

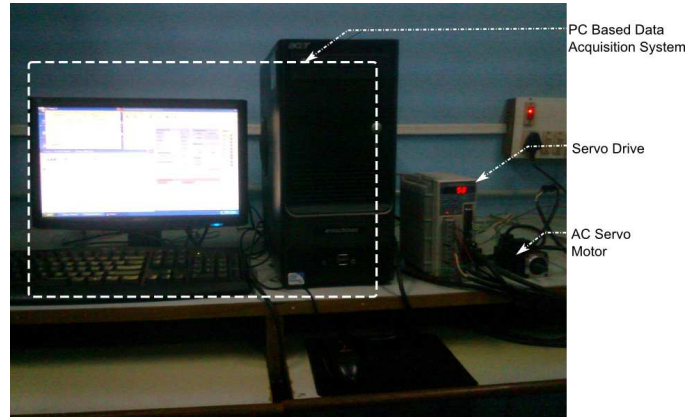

Figure 3.  Experimental set-up of the servo motor position control system.

In the servo drive there are three connectors viz. CN1, CN2 and CN3. CN1 is the input/output connector and is used to connect the external controller to the drive. This provides interfacing for analog speed and torque command signal, input pulse and reference voltage signal. CN2 is encoder connector and is used to connect integrated servomotor incremental encoder to drive input/output. CN3 is communication connector and is used to connect host controller via a serial communication cable. The drive used for experiment is comprised of Analog to Digital Converter (ADC), Digital to Analog Converter (DAC), control power, regenerative resistors, and protection circuit and display unit.

The Servo drive also has rectifier, dc link, inverter and Master Control Unit (MCU) that controls the speed, position and torque (current). In servo drive, rectification of ac signal is performed by converter and dc link. The dc link is used to remove the ripple from the converted signal. This rectified signal is then fed to the inverter to give controlled ac signal to drive the motor. Normally the encoder of servo motor provides feedback signal (i.e. position, speed and torque) to the position control, speed control and current control section. These control sections are monitored by MCU and results in the gating signal. This gating signal is used to drive the IGBT switch of the inverter and thus precise motion control of ac servo motor is attained. Advantech PCI-1220 Common Motion Driver has been used to control the servo system in the present case. Visual C++ code has been used for interfacing hardware with the control card to control the position of AC servo motor. The unit used for angular position, speed and acceleration is pulse per unit (ppu), ppu/s and ppu/s$^2$ respectively.

Data was collected for different velocity profiles with given set point i.e. with predefined position command. A customized C++ function was used to interface the PC based data acquisition system with the servo motor and record 5000 samples in 5 seconds. Data was gathered for two different sets of acceleration each associated with nine different velocities. As shown in Fig. 4-5, we get truncated ramp type position curve for an acceleration of 5×106 ppu/s², while for acceleration 1×106 ppu/s² we have an "S"-shaped curve. This change in the nature of position curve is due to fact that for high acceleration, the rate of increase in velocity is 5 times as compared to that at low acceleration. So the motor achieved its final velocity in much less time in case of operating at high acceleration and since integrating speed gives position so we get a truncated ramp in this case.

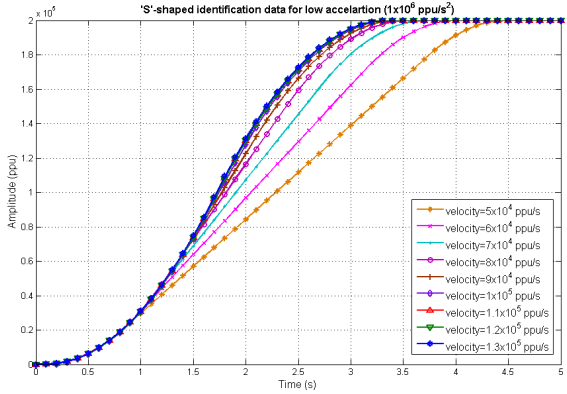

Figure 4. "S" shaped identification data for low acceleration.

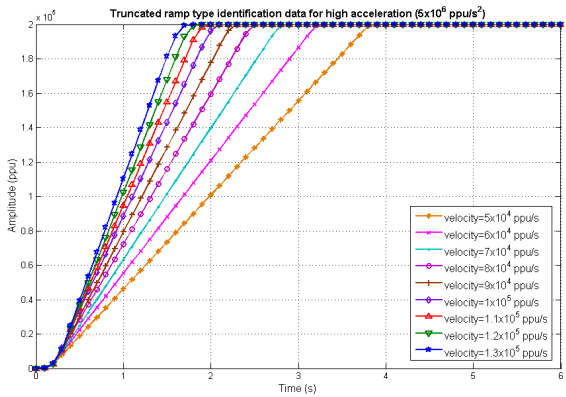

Figure 5. Truncated ramp type identification data for high acceleration.

Similar to the control rod to power model development of nuclear reactor, in this case a feed-forward multilayer ANN needs to be trained that maximally describe the nonlinear behavior of the ac servo position control system for high and low acceleration and different velocities, while producing the position as output. In both the target applications only the value of each sample could have been fed to train the neural networks. Since, dynamics of physical systems can not solely be described by the values at each sampling instants without the knowledge of its sampling time; the time elapsed (sampled) while recording the input-output system data has also been taken as another input to the multi-layer feed-forward ANN for improved identification/prediction performance.

## IV. RESULTS AND DISCUSSIONS

The Multiple-Input-Single-Output (MISO) data-set for both the nuclear reactor and the AC servo motor are then used for training of the two multilayer neural networks one for predicting the nonlinear system behavior of the reactor and the other for predicting the response of the ac servo motor from their local linear behaviors around different operating points. In both the cases, input and desired output values fed to the networks were normalized with respect to the maximum value, so as to ensure that all data used for training lie within the range of 0 to 1. The outputs of the trained neural networks were again scaled up to their original range to obtain the predicted response curves of the systems to compare with the actual system response curves.

TABLE I. TRAINING PERFORMANCE FOR VARIOUS ANN CONFIGURATIONS FOR THE IDENTIFICATION OF NUCLEAR REACTOR UNDER STEP-BACK FOR 20 INDEPENDENT RUNS

| Number of layers | Number of neurons in each hidden layer | Activation function | Mean of RMSE | Standard Deviation of RMSE |
|---|---|---|---|---|
| 1 | 5 | tansig | 0.0041 | 0.0028 |
| | | logsig | 0.0035 | 0.002 |
| | 10 | tansig | 0.0017 | 9.82×10⁻⁰⁴ |
| | | logsig | 0.0017 | 7.78×10⁻⁰⁴ |
| | 15 | *tansig* | *0.0014* | *8.87×10⁻⁰⁴* |
| | | logsig | 0.0014 | 9.19×10⁻⁰⁴ |
| | 20 | tansig | 0.0022 | 0.0012 |
| | | logsig | 0.0017 | 9.28×10⁻⁰⁴ |
| | 25 | tansig | 0.0020 | 7.35×10⁻⁰⁴ |
| | | logsig | 0.0021 | 0.0011 |
| 2 | 5 | tansig/tansig | 0.0031 | 0.0039 |
| | | tansig/logsig | 0.0093 | 0.0265 |
| | | logsig/tansig | 0.0031 | 0.0032 |
| | | logsig/logsig | 0.0036 | 0.0036 |
| | 10 | tansig/tansig | 0.0021 | 0.0016 |
| | | tansig/logsig | 0.0014 | 6.83×10⁻⁰⁴ |
| | | logsig/tansig | 0.0021 | 0.0012 |
| | | logsig/logsig | 0.0016 | 0.001 |
| | 15 | tansig/tansig | 0.003 | 9.94×10⁻⁰⁴ |
| | | tansig/logsig | 0.002 | 0.0012 |
| | | logsig/tansig | 0.0024 | 8.14×10⁻⁰⁴ |
| | | logsig/logsig | 0.0359 | 0.0963 |
| | 20 | tansig/tansig | 0.0034 | 0.0018 |
| | | tansig/logsig | 0.003 | 0.003 |
| | | logsig/tansig | 0.0035 | 0.0019 |
| | | logsig/logsig | 0.0027 | 8.08×10⁻⁰⁴ |
| | 25 | tansig/tansig | 0.0047 | 0.0029 |
| | | tansig/logsig | 0.0031 | 0.0024 |
| | | logsig/tansig | 0.0044 | 0.0019 |
| | | logsig/logsig | 0.0038 | 0.0032 |

Prediction was done by using a number of different neural network architectures differing in number of hidden layers and number of neurons in each layer. For each configuration the prediction was done for 20 independent runs. Tables I-II describe the root mean square values of errors of those network architectures and their standard deviation in predicting the system responses respectively. To check the consistency of each ANN configurations for capturing the system's dynamics as an input-output relationship, the RMSE of 20 independent runs has been chosen as the performance measure. This is justified from the fact that often large size multilayer ANN accurately establishes arbitrary nonlinear relation between any input-output data [7]-[8], but they might not show consistency in the mapping for different initial guesses of the Levenberg-Marquardt back-propagation algorithm since it is a gradient based optimization algorithm and thus often gets trapped in local minima. Hence there is always a trade-off between large size of the ANN structure and its average prediction accuracy.

TABLE II.  TRAINING PERFORMANCE FOR VARIOUS ANN CONFIGURATIONS FOR THE IDENTIFICATION OF AC SERVO-MOTOR POSITION CONTROL SYSTEM FOR 20 INDEPENDENT RUNS

| Number of layers | Number of neurons in each hidden layer | Activation function | Mean of RMSE | Standard Deviation of RMSE |
|---|---|---|---|---|
| 1 | 5 | tansig | 0.0218 | 0.007 |
| | | logsig | 0.0233 | 0.0081 |
| | 10 | tansig | 0.0106 | 0.0044 |
| | | logsig | 0.0117 | 0.0044 |
| | 15 | tansig | 0.0089 | 0.0049 |
| | | logsig | 0.0068 | 0.002 |
| | 20 | tansig | 0.0071 | 0.0026 |
| | | logsig | 0.0065 | 0.0021 |
| | 25 | tansig | 0.0056 | 0.0018 |
| | | logsig | 0.0056 | 0.0026 |
| 2 | 5 | tansig/tansig | 0.0207 | 0.0646 |
| | | tansig/logsig | 0.0132 | 0.0087 |
| | | logsig/tansig | 0.0093 | 0.0065 |
| | | logsig/logsig | 0.0116 | 0.0106 |
| | 10 | tansig/tansig | 0.0052 | 0.0107 |
| | | tansig/logsig | 0.0128 | 0.0437 |
| | | logsig/tansig | 0.0034 | 0.0025 |
| | | logsig/logsig | 0.0027 | 0.0012 |
| | 15 | tansig/tansig | 0.0054 | 0.0076 |
| | | tansig/logsig | 0.0027 | 0.0012 |
| | | *logsig/tansig* | *0.0021* | *0.0014* |
| | | logsig/logsig | 0.0029 | 0.0018 |
| | 20 | tansig/tansig | 0.0032 | 0.0022 |
| | | tansig/logsig | 0.0031 | 0.0021 |
| | | logsig/tansig | 0.0025 | 0.0015 |
| | | logsig/logsig | 0.0031 | 0.0021 |
| | 25 | tansig/tansig | 0.0033 | 0.0028 |
| | | tansig/logsig | 0.0026 | 0.0015 |
| | | logsig/tansig | 0.0041 | 0.0045 |
| | | logsig/logsig | 0.0026 | 0.0015 |

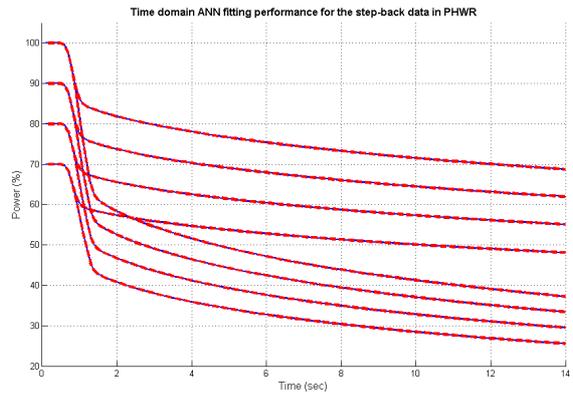

Figure 6. ANN fitting performance for nuclear reactor power level control.

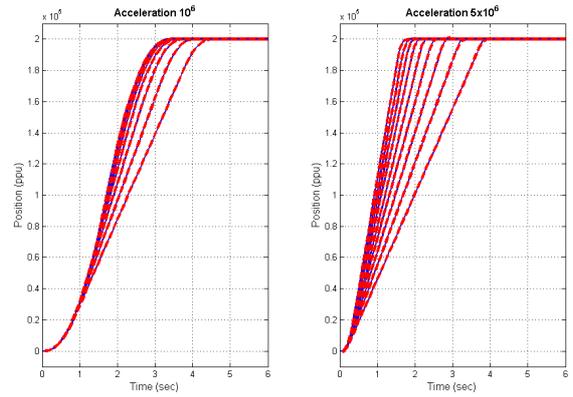

Figure 7. ANN fitting performance for AC servo position control system.

Figures 6-7 shows that the neural networks were able to faithfully predict the system responses of both the nuclear reactor and the AC servo motor around all the operating points. From Figures 8-9 it is quite clear that consistency of neural networks used is better for single hidden layer architecture than two hidden layer for both nuclear reactor and AC servo motor system which is also revealed from the statistical performance measures like mean and standard deviation of the RMSE etc.

From Table I it is evident that the neural network with a single hidden layer consisting of 15 neurons in that layer with hyperbolic tangent as the activation function performs best in predicting the system behavior for the nuclear reactor both in terms of mean and standard deviation of root mean square error for 20 independent runs. For the AC servo motor system (Table II), the neural network with two hidden layers consisting of 15 neurons in each layer with logarithmic sigmoid and hyperbolic sigmoid as the activation functions in the first and second hidden layer respectively predicts the system response most accurately in view of mean of root mean square error. Standard deviation for this particular network architecture is also quite small. Therefore, it can be concluded from the simulation results that for similar nonlinear system identification problems using the local linear behaviors, the optimum configuration of multi-layer feed-forward ANN, representing the system's original nonlinear dynamical behavior, should be judged using the average RMSE. Also, the respective optimum configuration depends on system's complexity and can't be chosen a priori.

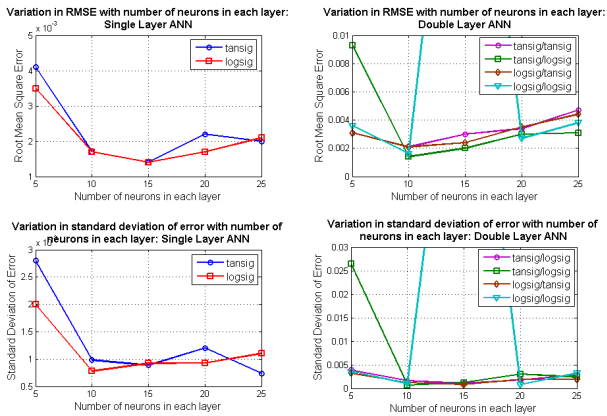

Figure 8. Statistical performance analysis of ANN based identification of nuclear reactor under step-back.

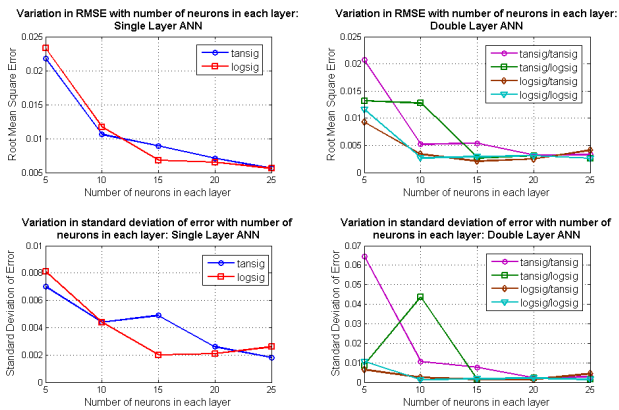

Figure 9. Statistical performance analysis of ANN based identification of AC servo position control system.

## V. CONCLUSION

A feed-forward neural network based system identification has been carried out from local linear information at different operating points for two practical non-linear systems i.e. a nuclear reactor and an AC servo position control system. Simulation results indicate that the proposed ANN based system identification methodology can successfully establish input-output mapping from the local linear data for such nonlinear systems. Rigorous parametric study has been done to find the best fit ANN architecture for the two target applications. This provides an insight into designing the best neural network architecture for such nonlinear systems and might provide directions to the operators utilizing nonlinear system identification for control or signal processing purposes.